\documentclass[12pt]{article}
\usepackage{a4wide,graphicx,amssymb}

\newcommand{\lmmt}[1]{\, \mathrm{L_{t\mu}^2}}

\begin{document}
\title{\vskip-3cm{\baselineskip14pt
}
\vskip.4cm
Three Loop Top Quark Contributions \\ to the $\rho$ Parameter 
\vskip.4cm
}
\author{
M.~Faisst, J.H.~K\"uhn, T.~Seidensticker, and O.~Veretin
\\[3em]
{\it Institut f\"ur Theoretische Teilchenphysik,} \\
{\it Universit\"at Karlsruhe, D-76128 Karlsruhe, Germany}
}
\date{}
\maketitle

\begin{abstract}
We present results for the three-loop top quark contributions to the
$\rho$ parameter in the limit of large top quark mass. The simultaneous
dependence on the mass of the Higgs boson $M_H$ and the mass of the top
quark $m_t$ is obtained from expansions in the range of $M_H$ around
$m_t$ and in the limit $M_H \gg m_t$. In combination with the previous
result for $M_H = 0$ the dependence of the $\rho$ parameter on the
leading Yukawa contributions, i.e.~on $m_t$ and $M_H$, is well under
control for all mass values of practical importance. 
The effects lead to a shift in the $W$ mass in the order of 5 MeV and 
are relevant for precision measurements at TESLA.
\end{abstract}
\section{Introduction}

The standard model predicts a strong influence of virtual heavy top
quarks in the low energy limit on four-fermion processes through their
virtual contributions to the W and Z boson self energies
\cite{rhoone}. This effect allows the indirect determination of the
mass of the Higgs boson and is generally essential for electroweak
precision tests. 

The bulk of the heavy top quark corrections to the W and Z self energies
can be collected in a deviation of the so called $\rho$ parameter from
the tree level result $\rho_{\mbox{\small tree}}=1$. The $\rho$
parameter is usually defined by the ratio of the neutral and charged
current coupling constants at zero momentum transfer \cite{rhoone}: 
\begin{equation}
\rho = \frac{J_{NC}(0)}{J_{CC}(0)} \,, 
\label{rho}
\end{equation}
where $J_{CC}(0)$ is given by the Fermi coupling constant $G_F$ determined
from the $\mu$-decay rate whereas $J_{NC}(0)$ is measured by neutrino
scattering on electrons or hadrons.

  In leading order contributions from the Yukawa coupling of the top quark
to $\rho$ are parameterized in terms of $\Delta\rho$. This contributions
stem from the transversal parts of the self energies of the
$W$- and $Z$-boson:
\begin{equation} \label{eq:def}
\rho = \frac{1}{1-\Delta\rho} \approx 
  \frac{1}{c^2} \, \frac{M_W^2 - \Pi_T^{WW}(0)}{M_Z^2 - \Pi_T^{ZZ}(0)}, 
\end{equation}
where $c^2=M_W^2/M_Z^2$ represents the cosine of the weak mixing angle,
defined in the standard on-shell scheme. Corrections from vertex and box
diagrams always involve extra powers of the weak coupling constant
$g_{weak}^2$ and are thus suppressed by powers of
$(g_{\rm weak}/g_{\rm Yukawa}^{\rm top})^2$, i.e.~$(M_W/m_t)^2$. 

In this context $\Pi_T^{WW}(0)$ and  $\Pi_T^{ZZ}(0)$ can be taken as the
bare one-particle irreducible amplitudes. The divergencies are absorbed
by the Higgs mass, top mass, and W boson mass renormalization, i.e.~by
expressing the bare $m_t$, $M_H$, and $M_W$ in terms of the renormalized
quantities. In the following we will first express the final answer in
terms of the $\overline{\mbox{MS}}$ top mass $m_t$, the on-shell Higgs
mass $M_H$, and the low energy  Fermi constant $G_F$ obtained from
$\mu$-decay. The latter is introduced to absorb the vector boson
mass $M_W$. In a last step the transformation from the
$\overline{\mbox{MS}}$ top quark mass $m_t$ to the on-shell top quark
mass $M_t$ will be performed.

The first two-loop results for $\Delta \rho$ in the limits $M_H
\rightarrow 0$ and  $M_H \gg m_t$ for large $m_t$ can be found in
\cite{rhotwo}, for arbitrary $M_H$ in \cite{rhofleischer}. The result
for $\Delta \rho$ of order $G_F m_t^2 \alpha_s^2$ was obtained in
\cite{tarasov}. By including terms of $\mathcal{O} ( M_W^2 / m_t^2 )$
this has even been extended to predictions for $\Delta r$ in three-loop
approximation.

The present paper is concerned with mixed QCD and electroweak
contributions of order $( G_F m_t^2 )^2 \alpha_s$ and purely electroweak
contributions of order $( G_F m_t^2 )^3$. It extends the result of
\cite{rhothree} where the special case $M_H=0$ has been
considered. Using the technique of heavy mass expansion for the limit
$M_H \gg m_t$ and considering sufficiently many terms leads to stable
predictions down to Higgs Mass values as low as twice $m_t$. $\Delta
\rho$ is furthermore evaluated for $M_H$ around $m_t$, and with the help
of expansions, a stable approximation is provided in the region from
close to zero up to twice $m_t$. Combining the results from the
different regions, the three-loop corrections to the $\rho$ parameter
are well under control.

An outline of the calculation, including a short discussion of the
renormalization, is given in section \ref{sec:diaren}. In section
\ref{sec:ms2os} the transformation of the top quark mass from the
$\overline{\mbox{MS}}$ scheme to the on-shell top mass is
treated in detail. Our results\footnote{Only numerical results are
  presented for the values of the coefficients of the various
  expansions. Their representations in terms of fractions and
  transcendental numbers can be obtained from the authors.}  at order
$G_F^3$ and $\alpha_s G_F^2$ are presented in sections \ref{sec:msbar}
and \ref{sec:os} in the $\overline{\mbox{MS}}$ and on-shell definition
for the top quark mass respectively.

\section{\label{sec:diaren}Treatment of diagrams and renormalization}

The computation of the three-loop contributions to $\Delta \rho$ at
orders $G_F^3$ and $\alpha_s G_F^2$ for $M_H \approx m_t$ and $M_H \gg
m_t$ is similar to the one discussed in \cite{rhothree} for $M_H =
0$. The same Feynman diagrams contribute: $W$ and $Z$ boson self
energies up to three loops, top quark self energies up to two loops, 
one-loop Higgs boson self energies, and Higgs tadpoles up to two loops.
All diagrams were generated using the program {\tt QGRAF} \cite{QGRAF}.

In comparison to \cite{rhothree} a non-vanishing Higgs mass introduces a
second mass scale into the problem. In order to calculate such diagrams
we consider two different ranges of the Higgs mass: First,
in the range of $M_H$ around $m_t$ we perform a straightforward Taylor
expansion in the mass difference $M_H - m_t$ around the point $M_H =
m_t$. In the second region, i.e.~for $M_H \gg m_t$, we use asymptotic
expansions \cite{asy} and made use of the program {\tt EXP}
\cite{EXP}. In both cases the resulting diagrams contained at most one
non-vanishing mass scale.

From equation (\ref{eq:def}) the vector boson self energies can be taken
at zero external momentum. After the application of the expansion
procedures the resulting integrals are of tadpole type with one mass
scale. Their computation can be performed using the program package {\tt
  MATAD} \cite{MATAD} written in {\tt FORM} \cite{FORM}. Apparently the
Higgs tadpoles, which are needed to renormalize the vacuum expectation
value of the Higgs boson, can be treated using the same program
package. To renormalize the mass of the Higgs boson, the computation of
on-shell Higgs boson self energy diagrams is necessary. The
corresponding renormalization constant is needed to one-loop order only
and the relevant integrals are straightforward. The remaining two-loop
top quark mass renormalization is, however, more involved, in particular 
in the on-shell scheme. Details on this issue are given in sections
\ref{sec:msbar} and \ref{sec:ms2os}.

The whole calculation was performed using dimensional regularization and
anticommuting $\gamma_5$. This prescription preserves the Ward--Takahashi
identities which relate the self energies of the gauge bosons to the
non-diagonal self energies of the gauge-to-goldstone and the goldstone
boson self energies. The validity of these identities was verified by
explicit calculations.

\section{\label{sec:msbar}The $\rho$-parameter in the
  $\overline{\mbox{MS}}$ scheme}

In this section we present our results for the $\overline{\mbox{MS}}$
definition of the top quark mass. The renormalization constant of the
top quark mass can be obtained by calculating the relevant one- and 
two-loop self energy diagrams in the limit of vanishing external
momentum. After the application of the expansions mentioned in section
\ref{sec:diaren}, the remaining integrals were computed using the
program package {\tt MATAD}.

The radiative corrections to $\Delta \rho$ depend on the
$\overline{\mbox{MS}}$ top mass $m_t \equiv m_t(m_t)$, the on-shell
Higgs mass $M_H$, the Fermi constant $G_F$ and the strong coupling
constant $\alpha_s$ and are written as follows: 
\begin{eqnarray} \label{eq:deltarho}
\Delta \rho &=& 
  x_t \, \Delta \rho^{(x_t)}     
+ x_t^2 \, \Delta \rho^{(x_t^2)} 
+ \frac{\alpha_s}{\pi} \, x_t \, \Delta \rho^{(\alpha_s x_t)}
+ \left( \frac{\alpha_s}{\pi} \right)^2 x_t \, \Delta
\rho^{(\alpha_s^2 x_t)}
\nonumber
\\&&\mbox{}
+ \frac{\alpha_s}{\pi} \, x_t^2 \, \Delta \rho^{(\alpha_s x_t^2)}
+ x_t^3 \, \Delta \rho^{(x_t^3)}
+ \cdots \, ,
\end{eqnarray}
with
\begin{displaymath}
x_t = \frac{G_F m_t^2}{8 \sqrt{2} \pi^2} \approx 3.0 \cdot 10^{-3}\,.
\end{displaymath}
Here and later on we always set
the $\overline{\mbox{MS}}$ renormalization scale $\mu$ equal to $m_t$

In this paper we evaluate the coefficients $\Delta \rho^{(x_t^3)}$ 
and $\Delta \rho^{(\alpha_s x_t^2)}$.
Numerically they  are given by:

$M_H = 0$:
\begin{eqnarray}
\Delta \rho^{(x_t^3)} &=& 47.73 \ , \\
\Delta \rho^{(\alpha_s x_t^2)} &=& 76.91 \ .
\end{eqnarray}

$M_H \approx m_t$: In this case we have computed five terms in the expansion
in $\delta$ defined by the relation $M_H = m_t ( 1 + \delta )$:
\begin{eqnarray} \label{eqn:ew_ms_eq}
\Delta \rho^{(x_t^3)} &=& -40.30 - 138.26\,\delta - 32.39\,\delta^2 +
10.32\,\delta^3 - 3.27\,\delta^4 - 0.01 \,\delta^5 +
\mathcal{O}(\delta^6) \, , \\ 
\Delta \rho^{(\alpha_s x_t^2)} &=&  93.07 + 21.27\,\delta +
0.53\,\delta^2 - 2.29\,\delta^3 + 1.14\,\delta^4-0.51\,\delta^5 +
\mathcal{O}(\delta^6) \, .
\label{eqn:as_ms_eq} 
\end{eqnarray}

\begin{figure}[!pt]
  \begin{center}
    \includegraphics[clip,width=14cm]{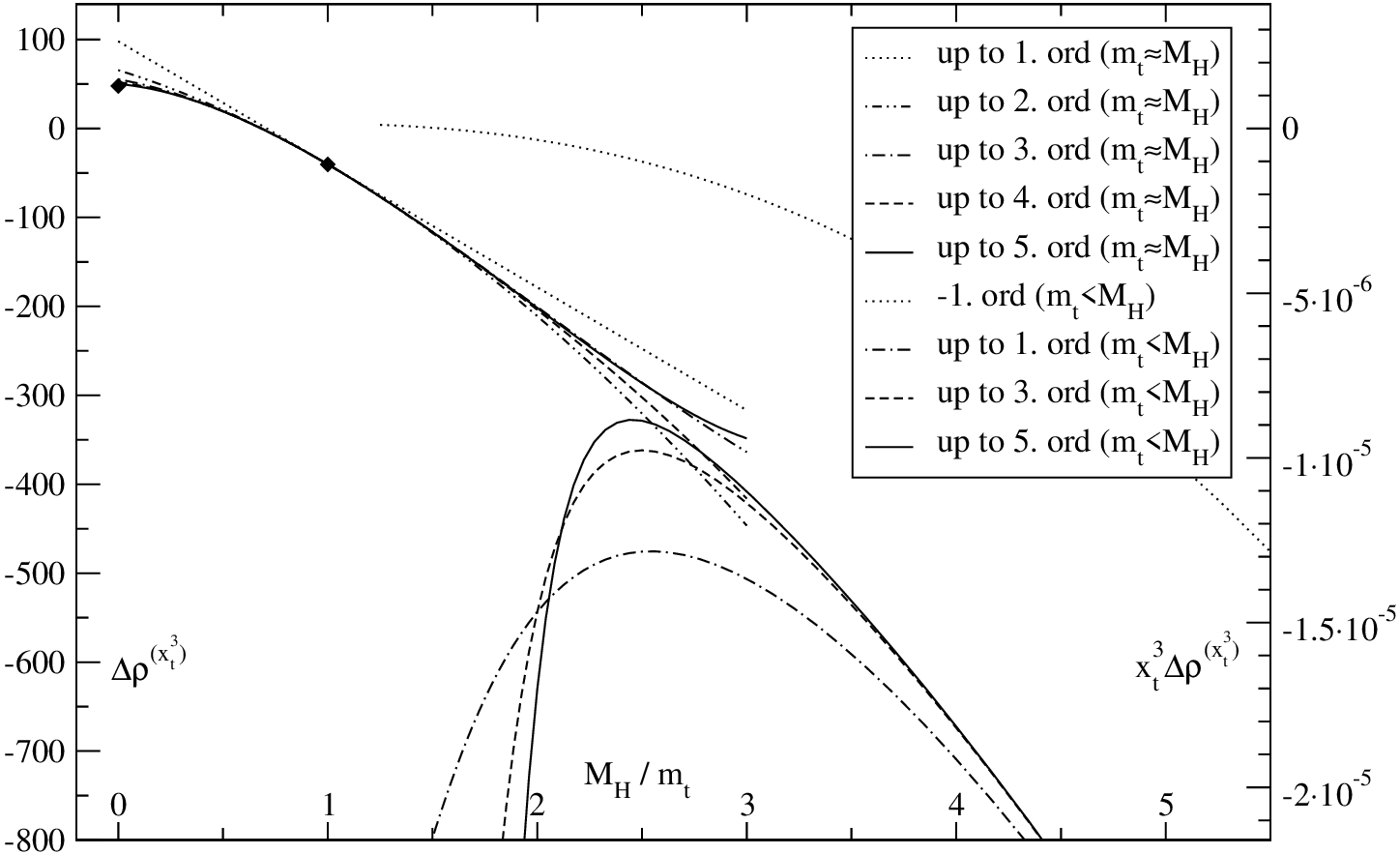}
  \end{center}
  \caption{\label{fig:ew_ms}Contributions of order $x_t^3$ to $\Delta
    \rho$ in the $\overline{\mbox{MS}}$ definition of the top quark
    mass. The black squares indicate the points where the exact result
    is known.} 
  \begin{center}
    \includegraphics[clip,width=14cm]{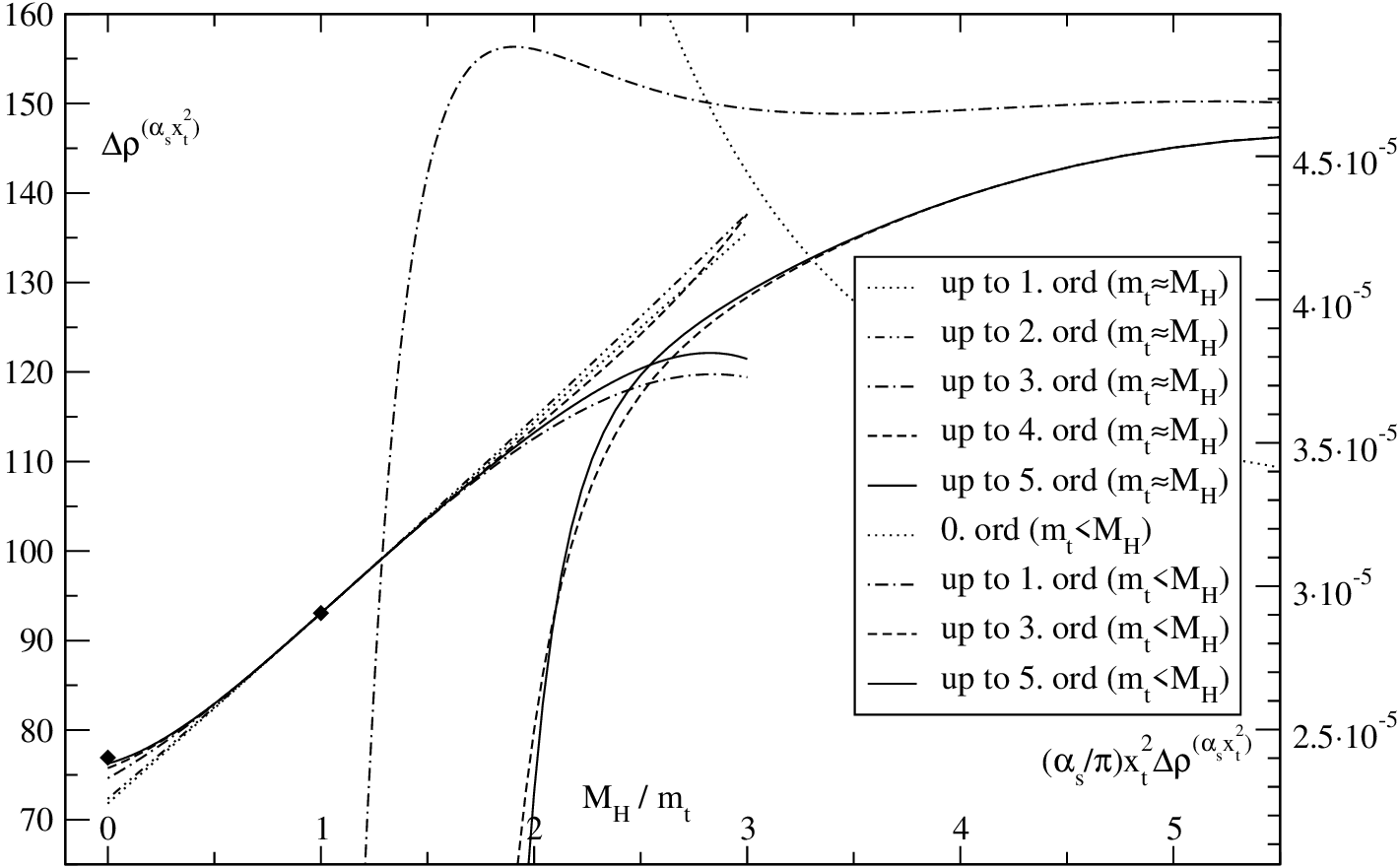}
  \end{center}
  \caption{\label{fig:as_ms}Contributions of order $\alpha_s x_t^2$ to
    $\Delta \rho$ in the $\overline{\mbox{MS}}$ definition of the
    top quark mass. The black squares indicate the points where the
    exact result is known.} 
\end{figure}
$M_H \gg m_t$: The application of the hard mass procedure leads to an
expansion in the ratio $y \equiv 4 m_t^2 / M_H^2$:
\begin{eqnarray}
\Delta \rho^{(x_t^3)}
 &=& \frac 1y ( -12.80 + 24.75\,\log y ) \nonumber \\
&-&192.86 - 394.05\,\log y - 275.02\,\log^2 y - 24.25\,\log^3 y 
\nonumber \\
&+& y\,\left( -337.88 + 408.59\,\log y - 233.35\,\log^2 y +
108.04\,\log^3 y \right) \nonumber \\
&+&  y^2\,\left( 82.56 - 378.31\,\log y + 26.14\,\log^2 y +
48.60\,\log^3 y \right) \nonumber \\ 
&+&  y^3\,\left( -82.58 - 291.03\,\log y - 11.37\,\log^2 y +
56.87\,\log^3 y \right) \\
&+&  y^4\,\left( -44.06 - 366.54\,\log y - 3.27\,\log^2 y +
59.62\,\log^3 y \right) \nonumber \\
&+&  y^5\,\left( -42.49 - 409.57\,\log y - 4.44\,\log^2 y +
64.70\,\log^3 y \right) \nonumber \\
&+& \mathcal{O}(y^6), \nonumber \\
\Delta \rho^{(\alpha_s x_t^2)} &=& 216.07 + 131.38\,\log y +
57.07\,\log^2 y + 9.00\,\log^3 y  \nonumber \\
&+&  y\,\left( -59.99 - 66.29\,\log y + 18.39\,\log^2 y -
19.22\,\log^3 y \right) \nonumber \\
&+&  y^2\,\left( -66.31 + 21.04\,\log y - 37.76\,\log^2 y -
5.57\,\log^3 y \right) \nonumber \\
&+&  y^3\,\left( -9.73 - 28.56\,\log y - 29.38\,\log^2 y -
5.46\,\log^3 y \right) \label{eqn:as_ms_mh} \\
&+&  y^4\,\left( -4.48 - 37.23\,\log y - 30.13\,\log^2 y -
5.25\,\log^3 y \right) \nonumber \\
&+&  y^5\,\left( -2.49 - 41.12\,\log y - 31.92\,\log^2 y -
5.14\,\log^3 y \right) \nonumber \\
&+& \mathcal{O}(y^6). \nonumber
\end{eqnarray}
In Fig. \ref{fig:ew_ms} and \ref{fig:as_ms} these results are plotted
as functions of the ratio of the Higgs mass $M_H$ and the top mass
$m_t$. The left scale gives the coefficient of the contribution under
consideration. The right scale includes the prefactor for $m_t = 165
\mbox{ GeV}$, so the full numerical effect on $\Delta \rho$ can be
directly read off. In Fig. \ref{fig:as_ms} the  numerical value
$\alpha_s(m_t) = 0.109$ was used for the right scale. 

The solid lines represent the sum of all terms displayed above. The
dashed curves represent successive approximations. From Figs.
\ref{fig:ew_ms} and \ref{fig:as_ms} we infer that the series given in
equations (\ref{eqn:ew_ms_eq}) to (\ref{eqn:as_ms_mh}) provides a
satisfactory approximation to the full result. In fact, a fairly smooth
transition between the two approximations is observed for $M_H \approx
2.5 m_t$. In addition the expansion seems to provide a good
approximation even down to $M_H=0$, where it can be compared to the
exact result.

Note, that the dominant term of the three-loop electroweak corrections
$\Delta \rho^{(x_t^3)}$ in the limit of large Higgs masses is
proportional to $M_H^2$. This is in agreement with the {\it screening
  theorem} \cite{veltact} stating that the highest power correction
depending on the mass of the Higgs boson is at least one power of
$M_H^2$ below the expectation based on naive power counting. The terms
$\sim M_H^2$ stem directly from the large mass expansion of the 
three-loop $W$ and $Z$ self energies. However, the absolute size is
small when the prefactor $x_t^3$ is included.

The smooth behavior of the two expansions in Figs. \ref{fig:ew_ms}
and \ref{fig:as_ms} together with the agreement with the result for
$M_H=0$ leads to the assumption that the three-loop corrections are well
under control in the $\overline{\mbox{MS}}$ scheme. 

\section{\label{sec:ms2os}Transition from $\overline{\mbox{MS}}$ to
  on-shell top quark mass}

In a final step the result for the on-shell definition of the
top quark mass are presented. This requires the ratio between the
$\overline{\mbox{MS}}$ renormalized top quark mass $m_t$ and the
on-shell mass $M_t$ in two-loop approximation, 
again in the limit of large top quark mass. 
Let us parameterize the ratio between $\overline{\mbox{MS}}$ in the following way
\begin{equation}
\label{massrelation}
\frac{m_t}{M_t} = 1 
  + \frac{\alpha_s}{\pi} C^{(\alpha_s)}
  + X_t C^{(X_t)}
  + \left(\frac{\alpha_s}{\pi}\right)^2 C^{(\alpha_s^2)}
  + \frac{\alpha_s}{\pi} X_t C^{(\alpha_s X_t)}
  + X_t^2 C^{(X_t^2)} + \dots \, ,
\end{equation}
using the expansion parameter
\begin{displaymath}
X_t = \frac{G_F M_t^2}{8 \sqrt{2} \pi^2} \approx 3.2 \cdot 10^{-3} \, ,
\end{displaymath}
which depends on the on-shell top quark mass $M_t$.

The results for $C^{(\alpha_s)}$, $C^{(X_t)}$ (e.g. \cite{Ca} and
\cite{Cxt} respectively), and $C^{(\alpha_s^2)}$ \cite{Ca2} 
are well known. Results for both $C^{(\alpha_s X_t)}$ and $C^{(X_t^2)}$
were obtained from the large mass and the mass difference expansion and
are needed for the present discussion.

The corresponding two-loop top quark self energy diagrams, already discussed
in section \ref{sec:msbar} for vanishing external momentum, have to be
recalculated for external momentum on mass shell. In general the
relevant diagrams can be divided into two classes: 
those containing at least one virtual Higgs boson and those without Higgs
bosons. The latter have already been evaluated for the computation of
$\Delta \rho$ in the case $M_H=0$ \cite{rhothree}. 
However, the former class of diagrams needs additional attention.

Let us first consider the limit $M_H \gg m_t$. Asymptotic expansions
reduce the two-loop on-shell integrals containing two mass scales to
combinations of one scale integrals. These one scale integrals include:
tadpole integrals up to two loops, one-loop on-shell integrals, and one
and two-loop integrals containing a small (compared to the internal
Higgs mass) external momentum. Each of these integrals is either
straightforward to calculate or can be treated using {\tt MATAD} and
{\tt MINCER}  \cite{MINCER}. Thus one obtains the on-shell top quark
self mass in terms of the $\overline{\mbox{MS}}$ mass in two-loop
approximation for the region $M_H \gg m_t$. The result reads as follows:
\begin{eqnarray}
C^{(X_t^2)}&=& \frac 1Y ( 1.61 - 18.00\,\log Y ) \nonumber \\
&+&  8.09 - 2.10\,\log Y - 1.50\,\log^2 Y \nonumber \\
&+&  Y\,\left( -4.50 + 6.05\,\log Y - 1.73\,\log^2 Y \right) \nonumber \\
&+&  Y^2\,\left( 14.60 - 9.74\,\log Y + 1.69\,\log^2 Y \right) \\
&+&  Y^3\,\left( 2.02 - 2.34\,\log Y + 0.64\,\log^2 Y \right) \nonumber \\
&+&  Y^4\,\left( 0.56 - 0.94\,\log Y + 0.32\,\log^2 Y \right) \nonumber \\
&+&  Y^5\,\left( 0.22 - 0.48\,\log Y + 0.17\,\log^2 Y \right) 
+ \mathcal{O}(Y^6) \, , \nonumber \\
C^{(\alpha_s X_t)}&=& 7.09 - 3.66\,\log Y + 1.50\,\log^2 Y \nonumber \\
&+&  Y\,\left( -3.78 + 2.93\,\log Y - 0.36\,\log^2 Y \right) \nonumber \\
&+&  Y^2\,\left( -0.89 + 0.99\,\log Y - 0.23\,\log^2 Y \right) \\
&+&  Y^3\,\left( -0.28 + 0.42\,\log Y - 0.15\,\log^2 Y \right) \nonumber \\
&+&  Y^4\,\left( -0.11 + 0.21\,\log Y - 0.11\,\log^2 Y \right) \nonumber \\
&+&  Y^5\,\left( 0.22 - 0.48\,\log Y + 0.17\,\log^2 Y \right)
+ \mathcal{O}(Y^6)\, , \nonumber
\end{eqnarray}
where $Y=4 M_t^2/M_H^2$ is used.

For $M_H$ around $m_t$ two-loop on-shell diagrams composed of several
massive and massless lines enter. Let us start with the case $M_H=m_t$
which involves two-loop on-shell diagrams with internal propagators
which are either massless or of mass $m_t$. The program packages
{\tt ONSHELL2} \cite{onshell} and {\tt TARCER} \cite{tarcer}, based on
the recurrence relations \cite{Tarasov:1997kx}, are specifically
tailored to this case. In the next step the Taylor expansion around this
point is performed and five terms of order $\alpha_s x_t$ and $x_t^2$
are evaluated. It should be mentioned, that in order $x_t^2$ there are
several diagrams with threshold at $M_H^2$. This means that in order to
expand these diagrams the threshold expansion should be applied
\cite{asy,Beneke:1997zp}. By explicit calculation, however, we find
that the first five coefficients are given by the naive Taylor
series. The results read as follows:
\begin{eqnarray}
C^{(X_t^2)}&=& 33.50 - 22.91\,\delta + 13.96\,\delta^2 -
2.30\,{\delta}^{3} + 2.71\,{\delta}^{4} - 3.46\,{\delta}^{5} +
\mathcal{O}(\delta^6) \, , \\
C^{(\alpha_s X_t)}&=& -22.15 + 34.76\,\delta - 6.83\,{\delta}^{2} +
2.64\,{\delta}^{3} - 1.32\,{\delta}^{4} + 0.19\,\delta^5 +
\mathcal{O}(\delta^6) \, , 
\end{eqnarray}
using $\delta$ defined by $M_H=M_t ( 1+\delta )$.

For completeness we also include the result for $M_H=0$:
\begin{equation}
C^{(X_t^2)}= 83.58 \ \mbox{ and } \ C^{(\alpha_s X_t)} = -17.53 \, .
\end{equation}
\begin{figure}[!pt]
  \begin{center}
    \includegraphics[clip,width=13.3cm]{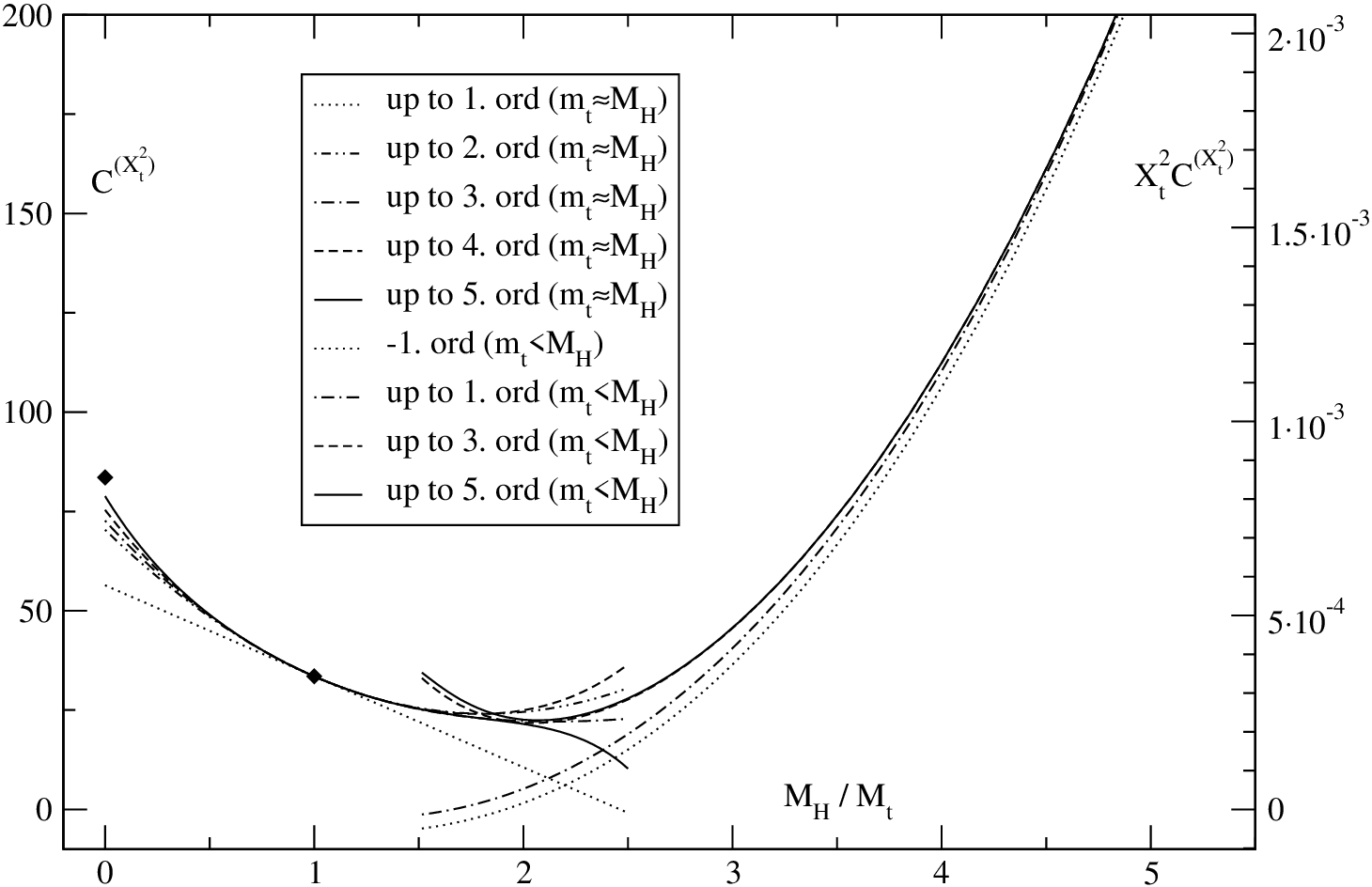}
  \end{center}
  \caption{\label{fig:ms2os_ew}Contributions of order $x_t^2$ to the
    relation between the $\overline{\mbox{MS}}$ and on-shell top quark
    mass. The black squares indicate the points where the exact result
    is known.} 
  \begin{center}
    \includegraphics[clip,width=13.3cm]{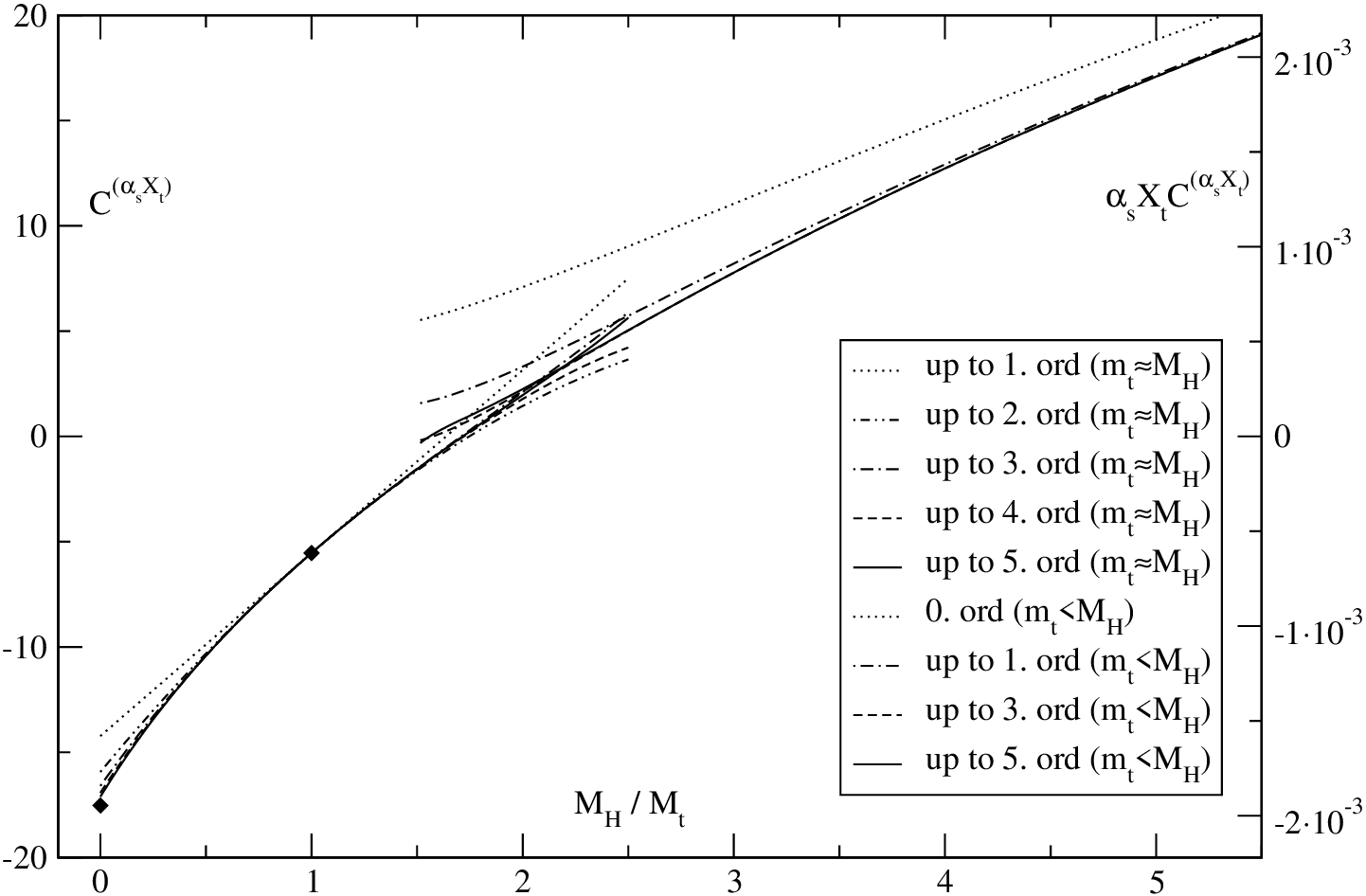}
  \end{center}
  \caption{\label{fig:ms2os_as}Contributions of order $\alpha_s x_t$ to
    the relation between the $\overline{\mbox{MS}}$ and on-shell top
    quark mass. The black squares indicate the points where the exact
    result is known.}
\end{figure}
The results are shown in Fig. \ref{fig:ms2os_ew} and \ref{fig:ms2os_as}.
A smooth transition between the two regions of large $M_H$ and
$M_H\approx M_t$ is observed. Furthermore, the expansion around
$M_H\approx M_t$, when extrapolated down to $M_H=0$, leads to a
remarkably good agreement with the value obtained from an analytical
calculation at $M_H=0$, a further and independent test of the
approximation procedure.

The result for the ratio between $\overline{\mbox{MS}}$ mass and
on-shell mass --- as well as the analytical expressions for the
coefficients given in sections \ref{sec:msbar} and \ref{sec:os} --- is
available from the authors upon request.

\section{\label{sec:os}$\Delta \rho$ expressed through the on-shell
  top quark mass}

Combining the results of sections \ref{sec:msbar} and \ref{sec:ms2os} we
compute $\Delta \rho$ in terms of the on-shell top quark mass $M_t$, the
on-shell Higgs mass $M_H$ and the low energy  Fermi constant $G_F$. Our
calculation reproduces the one- and two-loop results for the $\rho$
parameter at orders $G_F$, $G_F^2$ and $G_F \alpha_s$
\cite{rhoone,rhotwo,rhofleischer}.

The result for the on-shell definition of the top quark mass is
parameterized in the same way as the $\overline{\mbox{MS}}$ result in
section \ref{sec:msbar}.  In order to distinguish the two schemes we use
again the expansion parameter $X_t$, depending on the on-shell top quark
mass $M_t$.
\begin{figure}[!pt]
  \begin{center}
    \includegraphics[clip,width=14cm]{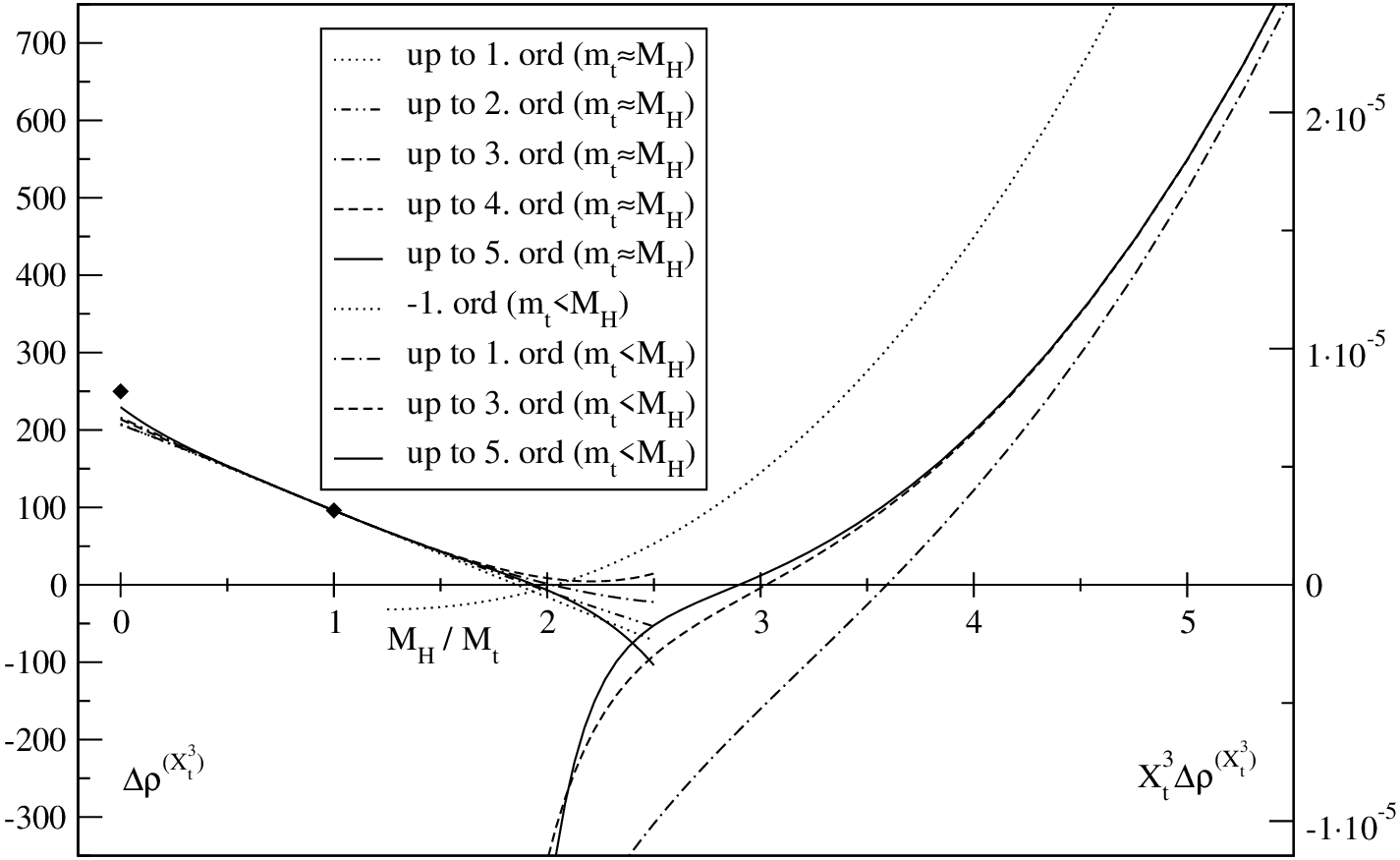}
  \end{center}
  \caption{\label{fig:ew_os}Contributions of order $X_t^3$ to
    $\Delta \rho$ in the on-shell definition of the top quark mass. The
    black squares indicate the points where the exact result is known.} 
  \begin{center}
    \includegraphics[clip,width=14cm]{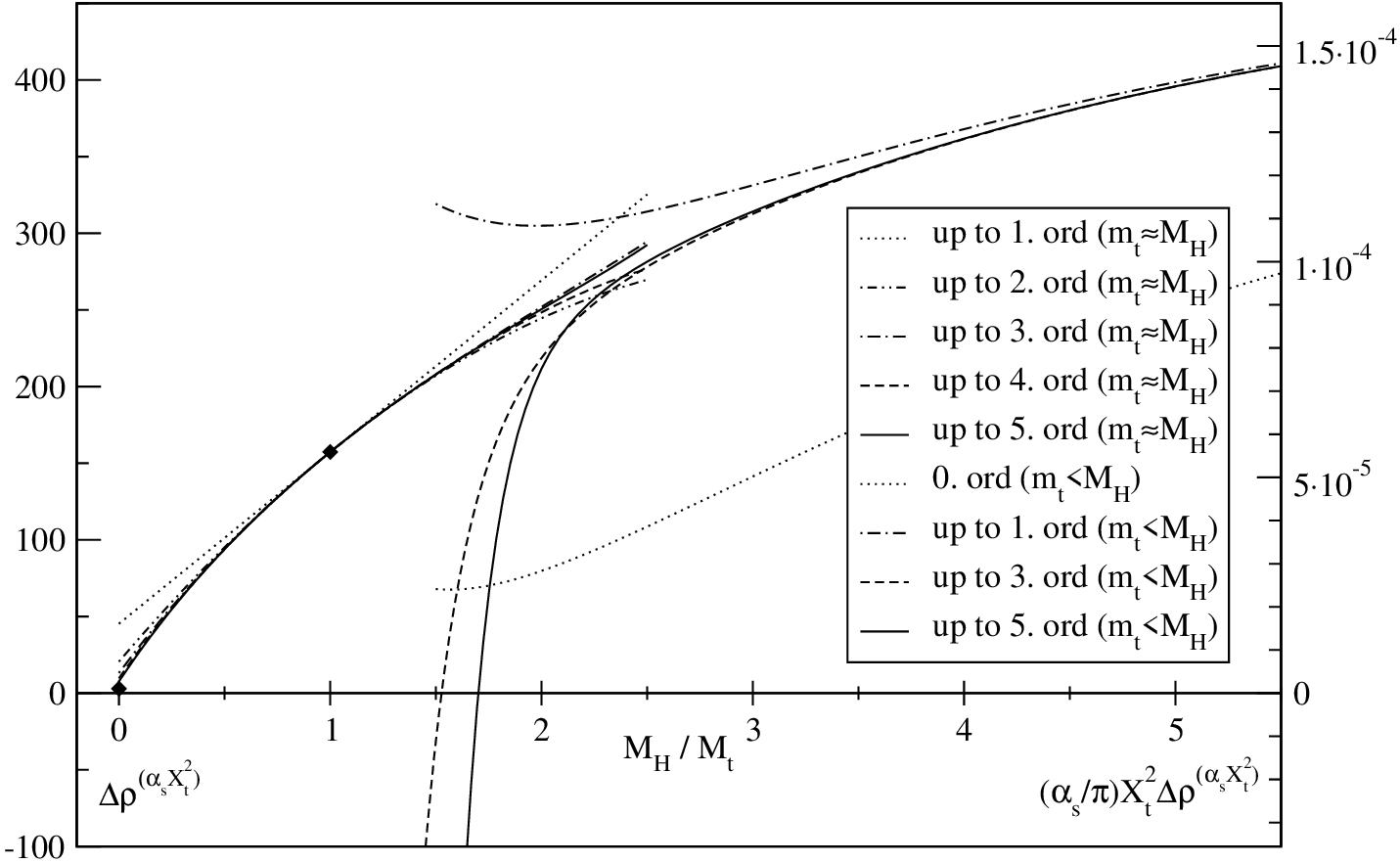}
  \end{center}
  \caption{\label{fig:as_os}Contributions of order $\alpha_s X_t^2$ to
    $\Delta \rho$ in the on-shell definition of the top quark mass. The
    black squares indicate the points where the exact result is known.}
\end{figure}

For completeness we also include the results for $M_H = 0$, computed in
\cite{rhothree}:
\begin{eqnarray}
\Delta \rho^{(X_t^3)} &=& 249.74 \ , \\
\Delta \rho^{(\alpha_s X_t^2)} &=& 2.94 \ .
\end{eqnarray}
For the expansion around $M_H=M_t$ we obtain:
\begin{eqnarray}
\Delta \rho^{(X_t^3)} &=& 95.92-111.98 \delta +8.099 \delta^2 + 9.36
\delta^3+7.27 \delta^4 + -15.60 \,\delta^5 + \mathcal{O}(\delta^6),\\
\Delta \rho^{(\alpha_s X_t^2)} &=& 157.295 + 112.00 \delta -24.73
\delta^2 +7.39 \delta^3 -3.52 \delta^4 + 2.06 \,\delta^5
+\mathcal{O}(\delta^6),
\end{eqnarray}
using $\delta$ defined by $M_H=M_t ( 1+\delta )$. 

In the limit $M_H \gg M_t$ we obtain
\begin{eqnarray}
\Delta \rho^{(X_t^3)} &=& 
\frac 1Y ( -3.17 - 83.25\,\log Y ) \nonumber \\
&-& 189.93 - 231.48\,\log Y - 142.06\,\log^2 Y 
+ 2.75\,\log^3 Y \nonumber \\
&+&  Y\,\left( -332.34 + 77.71\,\log Y - 68.67\,\log^2 Y +
51.79\,\log^3 Y \right)\nonumber \\
&+&  Y^2\,\left( 227.55 - 510.55\,\log Y + 87.77\,\log^2 Y +
6.41\,\log^3 Y \right) \nonumber \\
&+& Y^3\,\left( -58.40 - 329.18\,\log Y + 20.42\,\log^2 Y +
14.54\,\log^3 Y \right) \\
&+& Y^4\,\left( -36.14 - 381.88\,\log Y + 18.63\,\log^2 Y +
15.04\,\log^3 Y \right) \nonumber \\
&+& Y^5\,\left( -39.08 - 416.36\,\log Y + 13.76\,\log^2 Y +
17.19\,\log^3 Y \right) \nonumber \\
&+& \mathcal{O}(Y^6), \nonumber \\
\Delta \rho^{(\alpha_s X_t^2)} &=&
79.73 - 47.77\,\log Y + 42.07\,\log^2 Y + 9.00\,\log^3 Y
 \nonumber \\
&+&  Y\,\left( 225.16 - 179.74\,\log Y + 70.22\,\log^2 Y -
19.22\,\log^3 Y \right) \nonumber \\
&+&  Y^2\,\left( -76.07 + 25.33\,\log Y - 9.17\,\log^2 Y -
5.57\,\log^3 Y \right) \\
&+&  Y^3\,\left( -10.10 - 24.69\,\log Y - 0.30\,\log^2 Y -
5.46\,\log^3 Y \right) \nonumber \\
&+&  Y^4\,\left( -4.52 - 32.85\,\log Y + 0.72\,\log^2 Y -
5.25\,\log^3 Y \right) \nonumber \\
&+&  Y^5\,\left( -2.55 - 36.61\,\log Y + 1.06\,\log^2 Y -
5.14\,\log^3 Y \right) \nonumber \\
&+&\mathcal{O}(Y^6) \, ,\nonumber
\end{eqnarray}
with
\begin{equation}
Y \equiv \frac{4 M_t^2}{M_H^2} \, .
\end{equation}

The results in the on-shell definition of the top quark mass are shown
in Fig. \ref{fig:ew_os} and \ref{fig:as_os}. Again, the solid line
corresponds to the numerical values including all terms of our
expansion. The left and right scales give the size of the
coefficient and the full effect on $\Delta \rho$ respectively.

\begin{figure}[!ht]
  \begin{center}
    \includegraphics[clip,width=15cm]{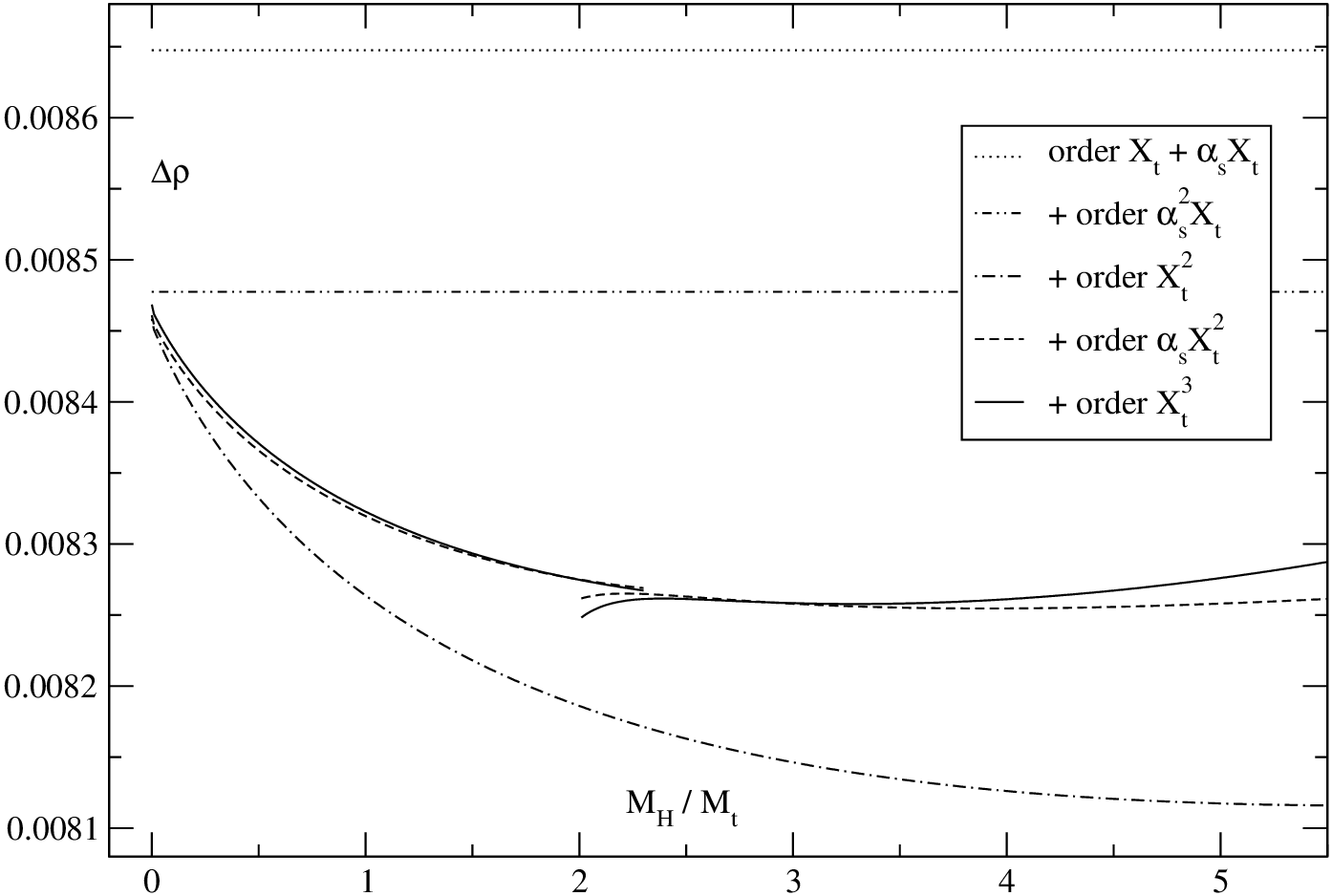}
  \end{center}
  \caption{\label{fig:deltarho}Effect of several contributions to $\Delta
    \rho$ in the on-shell definition of the top quark mass.}
\end{figure}

The $\alpha_s X_t^2$ contributions are sizeable and monotonously
increasing over the full range of the Higgs mass suggesting a smoothly
increasing behavior qualitatively similar to the well-known $X_t^2$
correction (displayed in Fig. \ref{fig:deltarho}). For $M_H=0$ the
result seems to be accidentally small, but increases rapidly for larger
Higgs mass values. The $X_t^3$ term exhibits a minimum in the range
between $M_H = 2 M_t$ and $M_H = 3 M_t$ and appears to be very small or
even negative in this region.

To compare the size of the three-loop contributions to the known one- and
two-loop terms we display their individual effects on $\Delta \rho$
in Fig. \ref{fig:deltarho}, including the corrections of order
$\alpha_s^2 X_t$ computed in \cite{tarasov}. We start with the sum of
the contributions of order $X_t$ and $\alpha_s X_t$ which are
independent of the Higgs mass and proceed by adding the particular
orders to $\Delta \rho$. To identify the individual contributions for
$M_H = 0$ we continued the value horizontally to the negative $M_H/M_t$
region. 

Let us first consider the $\alpha_s X_t^2$ and $X_t^3$ terms for the
case $M_H = 0$, discussed already in \cite{rhothree}. Inclusion of the
$\alpha_s X_t^2$ term leads to a change in $\Delta \rho$ of $1.0 \cdot
10^{-6}$, which is not visible in Fig. \ref{fig:deltarho}. The $X_t^3$
term increases $\Delta \rho$ by roughly $8.2 \cdot 10^{-6}$, again an
extremely small effect. The picture alters drastically for larger values
of $M_H$. The order $\alpha_s X_t^2$ correction is sizeable in the
region of $M_H \approx M_t$ and above. It amounts to approximately
$1/3$ of the two-loop contribution $\Delta \rho^{(X_t^2)}$, but appears
with the opposite sign. In contrast, the pure electroweak  three-loop
contribution $\Delta \rho^{(X_t^3)}$ yields an enhancement of about 1\%
compared to the order $X_t^2$ terms and is therefore negligible for
$M_H$ values of present interest.
\begin{figure}[t]
  \begin{center}
    \includegraphics[clip,width=15cm]{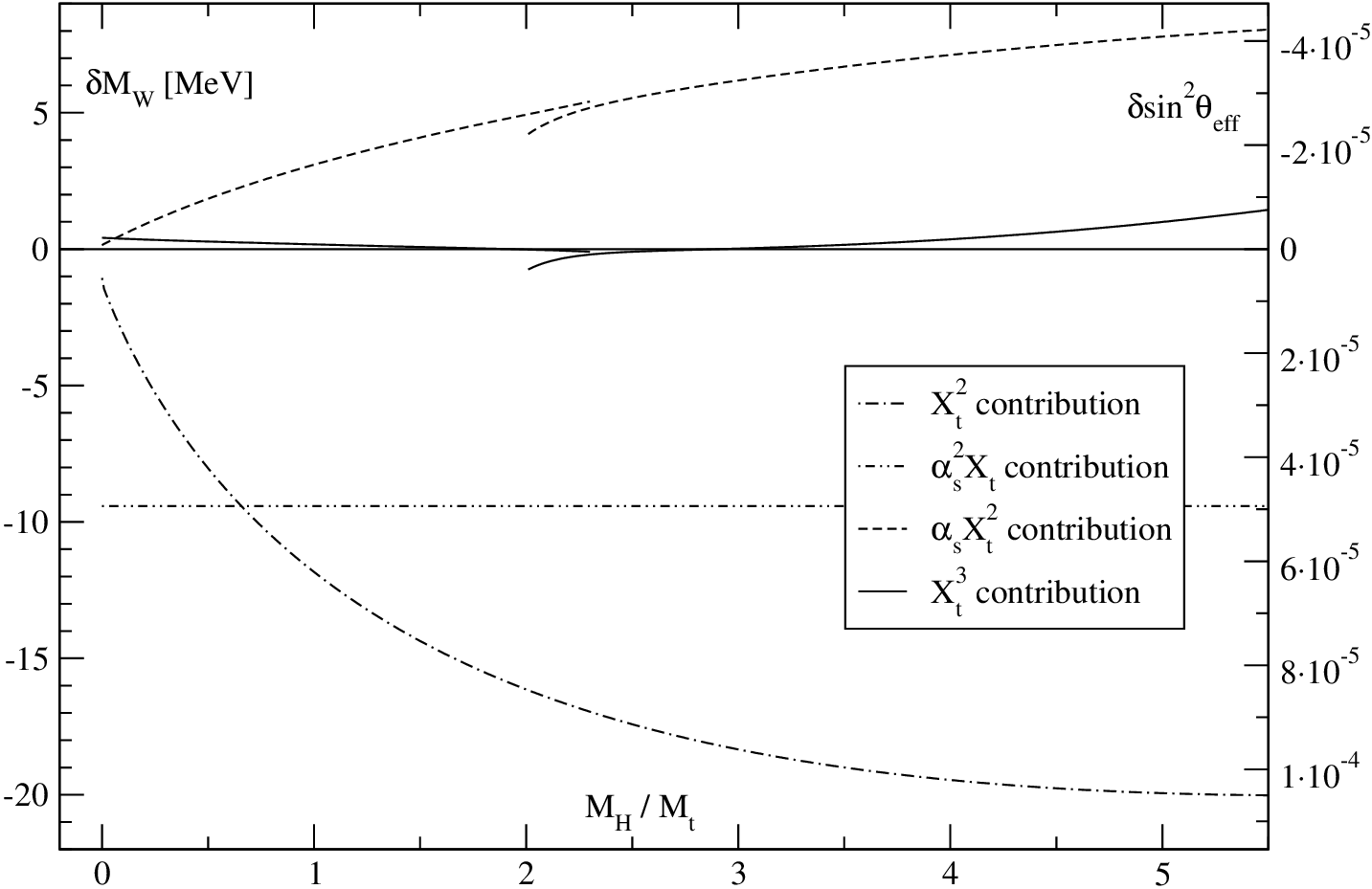}
  \end{center}
  \caption{\label{fig:shifts} Shifts in $M_W$ and in $\sin^2 \theta_W$
    from the various corrections to $\Delta \rho$.}
\end{figure}
These corrections to $\Delta \rho$ can be used to examine the resulting
shifts of observables like the $W$ mass and the effective weak mixing
angle, predicted for fixed values of $M_Z$, $\alpha$, and $G_F$ through
the equations
\begin{equation}
M_W^2 = \frac{\rho M_Z^2}{2} \left( 1 + \sqrt{1 - \frac{4 \pi
      \alpha}{\sqrt{2} \rho G_F M_Z^2 } \left(\frac 1{1-\Delta
        \alpha}+\dots \right)} \right)
\end{equation}
and
\begin{equation}
\sin^2 \theta_{\mbox{\small eff}} = 1-\frac{M_W^2}{\rho M_Z^2} = \frac
12 \left( 1 - \sqrt{1 - \frac{4 \pi \alpha}{\sqrt{2} \rho G_F M_Z^2 }
    \left(\frac 1{1-\Delta \alpha}+\dots \right)} \right),
\end{equation}
where $\Delta \alpha$ is the shift of $\alpha$ due to photon vacuum
polarization effects and the dots are nonleading remainder terms.
In Fig. \ref{fig:shifts} the resulting shifts are shown for the
different orders of $\Delta \rho$ in dependence of the Higgs mass (We
do not display the shift of order $\alpha_s X_t$ which amounts to
$-52.8$ MeV). This shift should be compared to $\delta M_W = 6 \mbox{
  MeV}$ and $\delta \sin^2 \theta_{\rm eff} = 1.3 \cdot 10^{-5}$ which
is the precision anticipated for TESLA operating at the $Z$-resonance
\cite{tesla}.

Alternatively we also present the shift in $M_t$ which would lead to a
comparable shift in $\Delta \rho$: 
\begin{equation}
\frac{\delta M_t}{M_t} \approx \frac 12 \, \frac {\delta \Delta \rho}{3
  X_t} \, .
\end{equation}
For $\delta \Delta \rho = 5 \cdot 10^{-5}$ this leads to a shift of
$m_t$ by about 500 MeV, which has to be compared to $\delta m_t \approx
\mathcal{O}(100 \mbox{MeV})$ anticipated for measurements at a future
linear collider.

For fixed $M_t$, $M_W$, and $M_Z$ our result for $\Delta \rho$ has to
be compensated by a shift in the prediction for $M_H$:
\begin{equation}
\frac{\delta M_H}{M_H} \approx \frac{12 \sqrt{2}}{11 G_F} \, \frac
{\cos^2 \theta_W}{\sin^2\theta_W} \, \delta\Delta\rho \approx 670 \;
\delta\Delta\rho
\end{equation}
For $M_H$ between 120 GeV and 200 GeV the term $\alpha_s X_t^2$ leads to
an upward shift between 4.5 and 7.5 GeV, the $X_t^3$ term to further
positive shift of less than 0.5 GeV.

The $X_t^2 \alpha_s$ term is, furthermore, 
comparable or larger than the non logarithmically enhanced light
fermion electroweak two-loop corrections \cite{dr_fermionic}
and significantly larger than the purely bosonic two-loop contribution.
The $X_t^3$ is comparable in size to the
bosonic two-loop effects evaluated in \cite{dr_bosonic}.

{\it Summary}: Combining the results from the extreme regions $M_H=0$,
$M_H \approx M_t$, and $M_H \gg M_t$ a prediction for the $X_t^3$ and
$\alpha_s X_t^2$ contributions to the $\rho$ parameter has been
obtained, which covers in a reliable way the full region of $M_H$. The
magnitude of these terms is comparable to the non-enhanced two-loop
corrections.

\section*{Acknowledgments} 
The authors would like to thank M.~Steinhauser who extended the program
package {\tt MATAD} to make this work possible. We are greatful to
K.G.~Chetyrkin and G.~Weiglein for useful discussions. We thank M.~Awramik
and M.~Czakon for tracing a misprint in an earlier version of this paper.

This work was supported by the {\it Graduiertenkolleg
``Hochenergiephysik und Teilchenastrophysik''}, by BMBF under grant
No. 05HT9VKB0, and the {\it DFG-Forschergruppe ``Quantenfeldtheorie,
  Computeralgebra und Monte-Carlo-Simulation''} (contract FOR 264/2-1).


\def\app#1#2#3{{\it Act.~Phys.~Pol.~}{\bf B #1} (#2) #3}
\def\apa#1#2#3{{\it Act.~Phys.~Austr.~}{\bf#1} (#2) #3}
\def\cmp#1#2#3{{\it Comm.~Math.~Phys.~}{\bf #1} (#2) #3}
\def\cpc#1#2#3{{\it Comp.~Phys.~Commun.~}{\bf #1} (#2) #3}
\def\epjc#1#2#3{{\it Eur.\ Phys.\ J.\ }{\bf C #1} (#2) #3}
\def\fortp#1#2#3{{\it Fortschr.~Phys.~}{\bf#1} (#2) #3}
\def\ijmpc#1#2#3{{\it Int.~J.~Mod.~Phys.~}{\bf C #1} (#2) #3}
\def\ijmpa#1#2#3{{\it Int.~J.~Mod.~Phys.~}{\bf A #1} (#2) #3}
\def\jcp#1#2#3{{\it J.~Comp.~Phys.~}{\bf #1} (#2) #3}
\def\jetp#1#2#3{{\it JETP~Lett.~}{\bf #1} (#2) #3}
\def\mpl#1#2#3{{\it Mod.~Phys.~Lett.~}{\bf A #1} (#2) #3}
\def\nima#1#2#3{{\it Nucl.~Inst.~Meth.~}{\bf A #1} (#2) #3}
\def\npb#1#2#3{{\it Nucl.~Phys.~}{\bf B #1} (#2) #3}
\def\nca#1#2#3{{\it Nuovo~Cim.~}{\bf #1A} (#2) #3}
\def\plb#1#2#3{{\it Phys.~Lett.~}{\bf B #1} (#2) #3}
\def\prc#1#2#3{{\it Phys.~Reports }{\bf #1} (#2) #3}
\def\prd#1#2#3{{\it Phys.~Rev.~}{\bf D #1} (#2) #3}
\def\pR#1#2#3{{\it Phys.~Rev.~}{\bf #1} (#2) #3}
\def\prl#1#2#3{{\it Phys.~Rev.~Lett.~}{\bf #1} (#2) #3}
\def\pr#1#2#3{{\it Phys.~Reports }{\bf #1} (#2) #3}
\def\ptp#1#2#3{{\it Prog.~Theor.~Phys.~}{\bf #1} (#2) #3}
\def\sovnp#1#2#3{{\it Sov.~J.~Nucl.~Phys.~}{\bf #1} (#2) #3}
\def\tmf#1#2#3{{\it Teor.~Mat.~Fiz.~}{\bf #1} (#2) #3}
\def\yadfiz#1#2#3{{\it Yad.~Fiz.~}{\bf #1} (#2) #3}
\def\zpc#1#2#3{{\it Z.~Phys.~}{\bf C #1} (#2) #3}
\def\ppnp#1#2#3{{\it Prog.~Part.~Nucl.~Phys.~}{\bf #1} (#2) #3}
\def\ibid#1#2#3{{ibid.~}{\bf #1} (#2) #3}
\def\jhep#1#2#3{{\it JHEP~}{\bf #1} (#2) #3}

\end{document}